# High-dimensional point forecast combinations for emergency department demand

Peihong Guo[1,*], Wen Ye Loh[1], Kenwin Maung[2], Esther Li Wen Choo[1], Borame Lee Dickens[3], Kelvin Bryan Tan[4], John Abishgenadan[1,5], Pei Ma[3], Jue Tao Lim[1]

[1]Lee Kong Chian School of Medicine, Nanyang Technological University, Singapore

[2]Department of Economics, Rutgers University, United States

[3]Saw Swee Hock School of Public Health, National University of Singapore, Singapore

[4]Ministry of Health, Singapore

[5]Tan Tock Seng Hospital, Singapore

**Abstract**

Current work on forecasting emergency department (ED) admissions focuses on disease aggregates or singular disease types. However, given differences in the dynamics of individual diseases, it is unlikely that any single forecasting model would accurately account for each disease and for all time, leading to significant forecast model uncertainty. Yet, forecasting models for ED admissions to-date do not explore the utility of forecast combinations to improve forecast accuracy and stability. It is also unknown whether improvements in forecast accuracy can be yield from (**1**) incorporating a large number of environmental and anthropogenic covariates or (**2**) forecasting total ED causes by aggregating cause-specific ED forecasts. To address this gap, we propose high-dimensional forecast combination schemes to combine a large number of forecasting individual models for forecasting cause-specific ED admissions over multiple causes and forecast horizons. We use time series data of ED admissions with an extensive set of explanatory lagged variables at the national level, including meteorological/ambient air pollutant variables and ED admissions of all 16 causes studied. We show that the simple forecast combinations yield forecast accuracies of around 3.81%–23.54% across causes. Furthermore, forecast combinations outperform individual forecasting models, in more than 50% of scenarios (across all ED admission categories and horizons) in a statistically significant manner. Inclusion of high-dimensional covariates and aggregating cause-specific forecasts to provide all-cause ED forecasts provided modest improvements in forecast accuracy. Forecasting cause-specific ED admissions can provide fine-scale forward guidance on resource optimization and pandemic preparedness and forecast combinations can be used to hedge against model uncertainty when forecasting across a wide range of admission categories.

Keywords: Forecasting; Forecast Combinations; Emergency Department; Machine Learning; High-dimensional Time Series Data

---

[*] Corresponding author

E-mail address: peihong001@e.ntu.edu.sg (P. Guo), lohwenye@gmail.com (W.Y. Loh), gm828@economics.rutgers.edu (K. Maung), L230001@e.ntu.edu.sg (E.L.W. Choo), ephdbsl@nus.edu.sg (B.L. Dickens), Kelvin_Bryan_TAN@moh.gov.sg (K.B. Tan), john_abisheganaden@ttsh.com.sg (J.Abisheganaden), mapei@nus.edu.sg (P. Ma), juetao.lim@ntu.edu.sg (J.T. Lim)

**Introduction**

Over the past 20 years, the volume of published literature on health forecasting has increased approximately five folds[1], driven primarily by greater availability of big data[2], wider diversity of disease surveillance data sources[3], as well as COVID-19 – which renews interest in pre-empting disease outbreaks[4]. Improving health forecasts remains a central priority for pandemic preparedness and optimization of resources during both inter-epidemic and epidemic phases of disease transmission. Health forecasts can help inform public health response, hospital resource planning and manpower allocation. Accurately ascertaining when surges in demand occur, and for which disease, can pre-empt appropriate institutional and public health measures to reduce strain on primary care resources[5]. In particular, decisions about hospital staffing, resource allocation, timing of public health communication messages, and the implementation of interventions designed to disrupt disease progression or transmission can be informed by forecasts.

A large number of forecasting models have been developed to conduct forecasts of emergency department (ED) load and peak load timings[6]. These are motivated by ED overcrowding which is recognized as a severe global health issue[7]. Overcrowding results from the mismatch between demand and supply of emergency medical services within the ED. ED overcrowding can lead to extended waiting times and treatment delays[8]. These prolonged delays can compromise the quality of care and elevate the risk of adverse outcomes for critically ill patients[9]. As ED load is situational and variable unlike hospital clinics where appointments can be scheduled in advance, various attempts have been made to pre-empt demands ahead of time. From traditional methods based on autoregressive or exponential algorithms to modern deep learning approaches[10], a wide variety of time series models have been developed for forecasting ED demand. Autoregressive integrated moving average (ARIMA) or variants have already been used in many ED forecasting studies and have dominated the field for a long time[11–13]. However, due to the complexity and stochastic nature of emergency department processes[14], traditional statistical methods often yield inaccurate forecasts[15]. Machine learning models have demonstrated superior performance in emergency department load forecasting. For example, a study that forecasts ED occupancy 24-hour ahead shows that DeepAR, LightGBM, etc. can outperform traditional benchmarks, with up to 15% improvement[16]. Multiple deep learning models have also been evaluated and a forecasting model utilising the Variational AutoEncoder (VAE) has demonstrated better performance versus alternatives[17]. Another study shows that the attention model is superior to the LSTM model, and the best results are achieved after incorporating calendar data[18].

However, to date, it is unclear whether improvements in forecast accuracy in forecasting ED admissions can be yield from (**1**) forecast combinations – especially with a large number of forecasting models (**2**) incorporating of a large number of environmental and anthropogenic covariates which have been found to be associated to ED risk and (**3**) forecasting total ED causes by aggregating cause-specific ED forecasts. Most multi-model forecast comparisons for ED admissions focus on a singular disease type and individual models which severely limits its applicability to finer-scale resource optimization[19]. In the United States alone, around 139.8 million ED visits occur in 2021, and the cause of admission could be due to a large number of etiologies across both chronic and infectious disease health domains[20] [21]. Admissions could be influenced by social, environmental and anthropogenic factors, such as socioeconomic deprivation[22], ambient air quality[23] and primary care referrals[24]. Proposing a singular ED load forecasting model and ignoring associated covariates would lack the required generalizability and robustness to provide accurate forecasts[25], especially over multiple disease categories and long study periods.

Therefore, in this study, we sought to address these research gaps in forecasting ED admissions by proposing to combine a large number of point forecasts to provide quick and robust forward guidance of cause-specific ED admissions. We hypothesize that forecast combinations can serve as an effective hedge against model uncertainty (i.e. which model works best for which disease) and unknown changes in disease dynamics by averaging out the idiosyncratic forecast errors of each model, thereby providing better forecasts against singular forecasting models. We conduct extensive forecasting exercises to examine the differences in forecast performance when including or excluding exogenous factors in forecasting models. Lastly, we examine whether aggregation of cause-specific ED forecasts to generate forward forecasts of total ED admissions improves upon models which only forecast total ED admissions.

To do so, we first collate a large number of environmental and anthropogenic covariates to train forecasting submodels to generate individual forward point forecasts. We then develop weighting schemes to aggregate point forecasts by using either simple combination schemes or combining them using constrained, high-dimensional supervised learning capable of accounting for point forecasts from many submodels, while inducing sparsity. Among many supervised learning schemes, we also explore regression averaging combinations based on random subset selection and random projection. We then compare these forecast combinations against alternatives. We provide a detailed analysis of the generalizability of our findings over multiple years, emergency department admission causes, and forecast horizons. Also, we identify key advantages and bottlenecks which enhance and limit the utility of combining forecasts respectively, using weekly emergency department admissions across multiple defined disease categories from 2009 to 2018. Our work shows overall strong evidence that forecast combinations which appropriately account for a large number of forecasting submodels can consistently outperform baseline models, as well as individual submodels, for a large number of ED admission causes. We also show that the inclusion of exogenous covariates can improve ED admission forecasts, and aggregation of cause-specific ED forecasts to generate forward forecasts of total ED admissions improves forecast accuracy for total ED admissions.

**Result**

Models with and without environmental covariates in the input have different forecast performances for total ED admissions. The inclusion of exogenous variables leads to decreased forecast errors for XGBoost, lightGBM and LASSO. The errors of RF and KNN do not change significantly. The other regularization methods even have higher errors, especially for longer horizons (Fig. 1A, 1B). Individual models show significantly different forecasting performances in various causes. These models perform well for some diseases but poorly for some others (Fig. D1–D16). We find that forecasting total ED admissions based on aggregating cause-specific ED forecasts reduced the forecast error in all forecast horizons, versus forecasts generated from a forecasting model trained on the outcome of all-cause ED admissions (Fig. 1B, 1C).

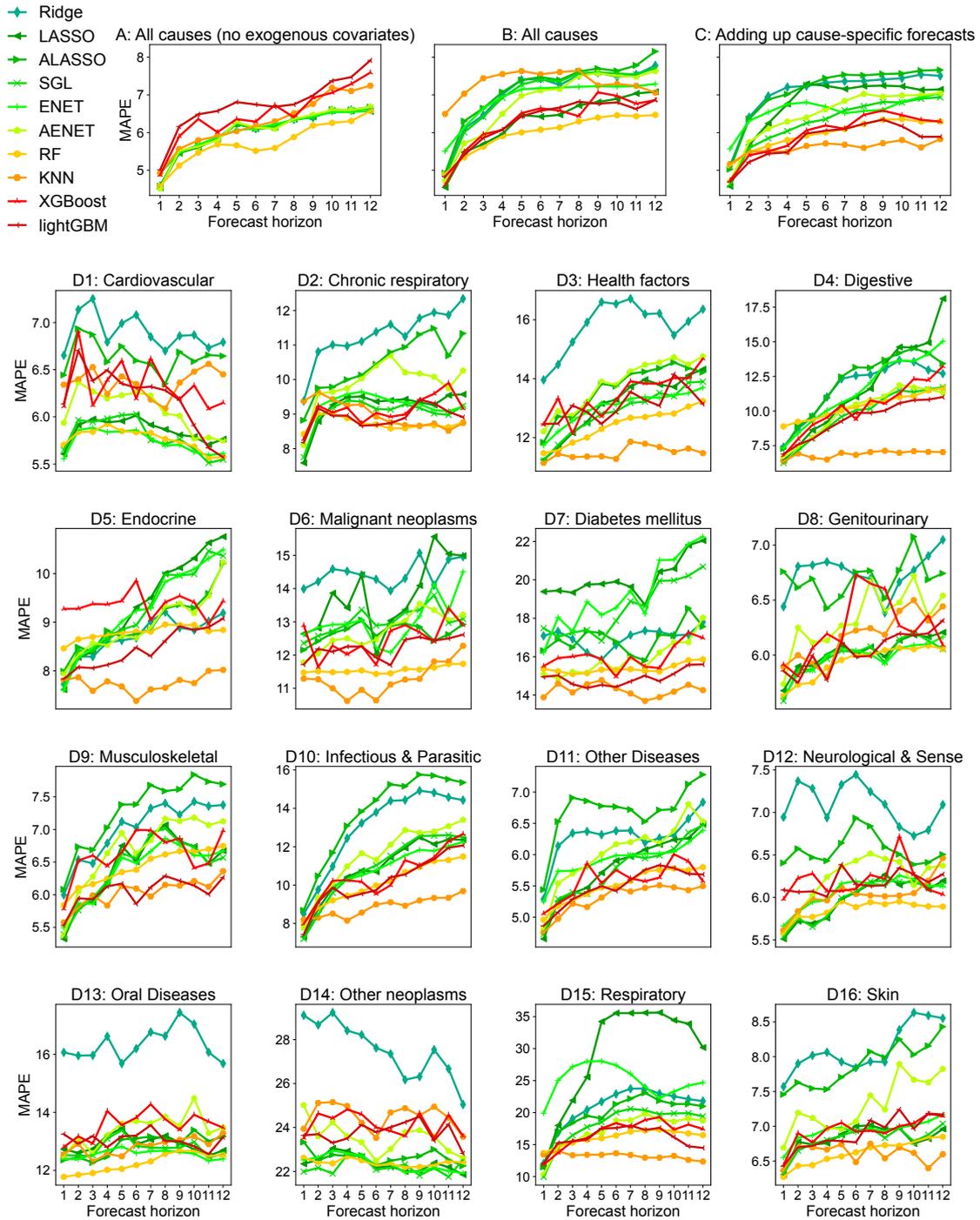

Fig. 1. Forecast assessment statistics for the all-cause and cause-specific forecasting models in the full forecast period. Mean absolute percentage forecast error (MAPE) of the 10 individual simple forecasting models which include: ridge regression (Ridge), Least Absolute Shrinkage and Selection Operator (LASSO), adaptive LASSO (ALASSO), sparse group LASSO (SGL), elastic net (ENET), adaptive elastic net (AENET), random forest (RF), k-nearest neighbors (KNN), XGBoost and lightGBMs, across the forecast horizon of 1-12 weeks ahead. (**A**) value assessed: forecasts of the total number of ED admissions for all causes; model inputs: autoregressive term (**B**) value assessed: forecasts of the total number of ED admissions for all causes; model inputs: autoregressive term and environmental covariates. (**C**) value assessed: sum of forecasts of the casue-specific ED admissions; model inputs: autoregressive term and environmental covariates. (**D1–D16**) value assessed: forecasts of the casue-specific ED admissions; model inputs: autoregressive term and environmental covariates.

In general, all 16 individual forecasting models generate different forecast errors across all 16 ED admission categories for all forecast horizons ahead (Fig. 2). Except for six causes (Fig. 2A6, 2B6, 2B7, 2D5, 2D6, 2D7), the forecasts of the forecast set of most models are better than those of the training set of the naïve method, especially for longer forecast horizons, as indicated by MASEs which are lower than 1. Pure autoregression AR(A) fitted with ordinary least square (OLS) consistently outperform the two autoregressive models with exogenous predictors. The nonparametric models all perform well, especially using KNN. The three tree-based ensemble models consistently perform similarly. The pure factor model transforms the exogenous information into a lower dimensional space, and thus avoiding overfitting. Its forecasting performance outperforms AR(A) in most ED categories, especially for respiratory infections. Several regularization strategies do not achieve expected gains in forecasting performance, especially the adaptive models. Adaptive LASSO is inferior to sparse group LASSO and LASSO in most disease categories (except diabetes mellitus, endocrine disorders, and respiratory infections), and adaptive elastic net is also inferior to elastic net in most categories. In addition, there is a tendency for the forecasts to mean-revert at longer forecast horizons, which leads to poorer capture of the peaks and troughs for each epidemiological week's admission numbers, and consequentially, a small decline in forecast performance at longer forecast horizons (Fig. 2, See S2 file for plots of case counts versus forecasts).

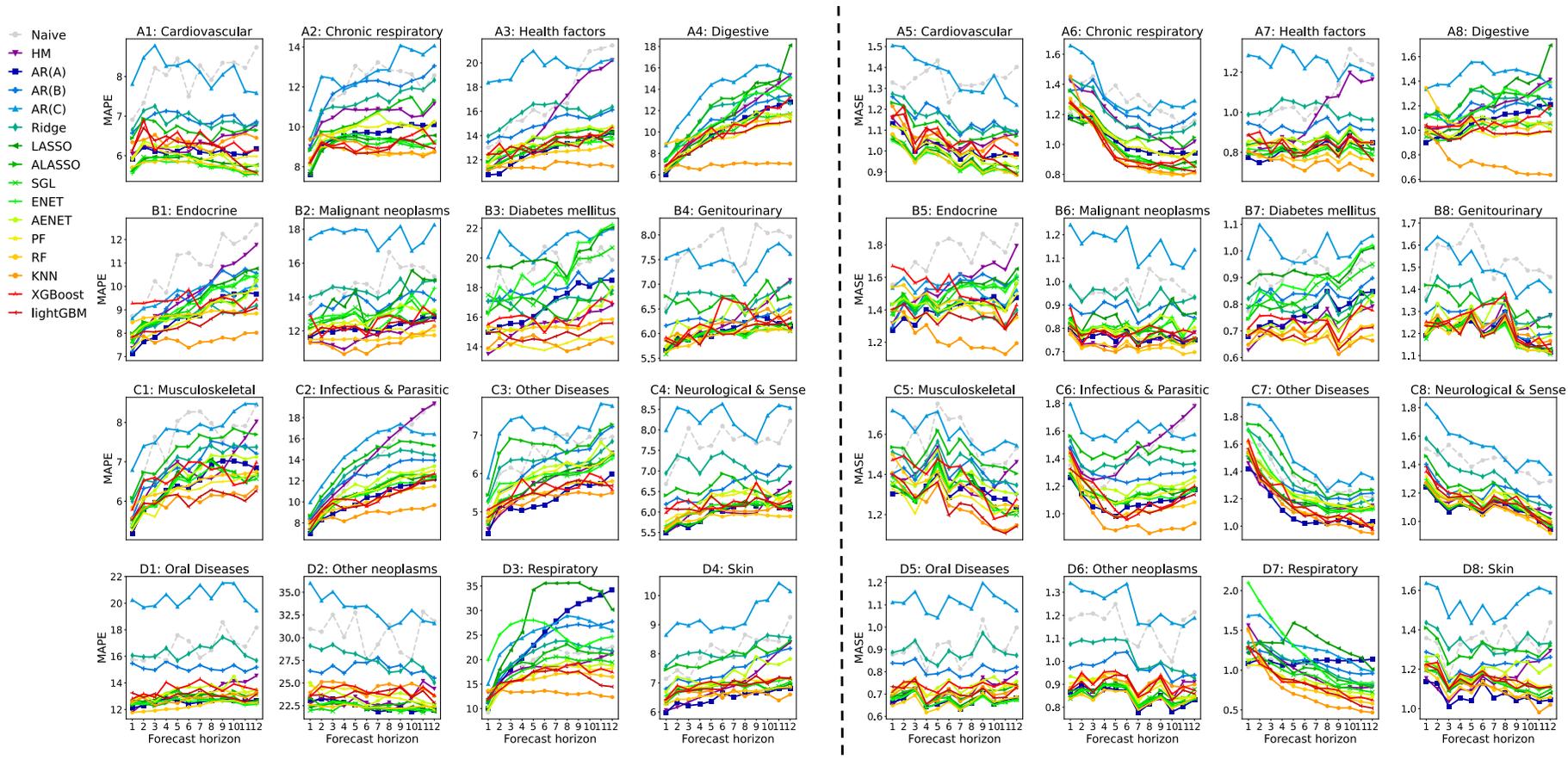

Fig. 2. Forecast assessment statistics in the full forecast period. Mean absolute percentage forecast error (MAPE) (left of the dotted line) and mean absolute scaled error (MASE) rates (right of the dotted line) of the 16 individual simple forecasting models which include 1 naïve model, 15 simple submodels: historical mean (HM), autoregression without exogenous variables [AR(A)], autoregression only with environmental variables [AR(B)], autoregression with all exogenous variables [AR(C)], ridge regression (Ridge), Least Absolute Shrinkage and Selection Operator (LASSO), adaptive LASSO (ALASSO), sparse group LASSO (SGL), elastic net (ENET), adaptive elastic net (AENET), pure factor model (PF), random forest (RF), k-nearest neighbors (KNN), 2 gradient boosting machines: XGBoost and lightGBMs, across the forecast horizon of 1-12 weeks ahead for individual disease categories.

Simple forecast combinations have the lowest error statistic for most disease categories across all forecast horizons (Fig. 3), indicating that the simple combinations schemes are able to improve on submodels for the majority of disease categories for long-term forecasting. In addition, starting from the 4-week forecasting window, supervised learning combinations have the lowest error statistic for some out of 16 disease categories (Fig. 3D–3L), which indicates that supervised forecast combinations outperform both simple submodels and simple combination schemes at a longer forecast horizon for some specific causes (mainly musculoskeletal, infectious and parasitic, and respiratory infection).

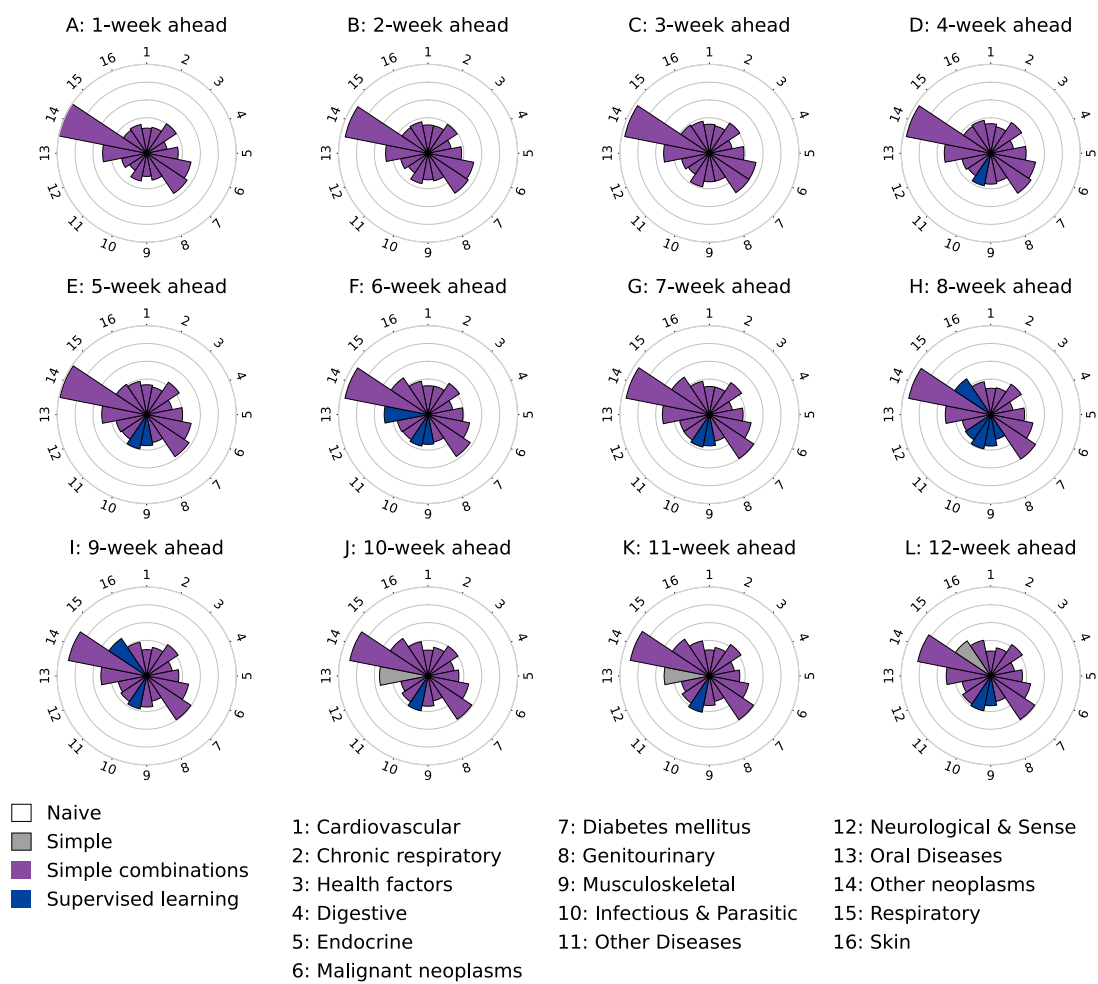

Fig. 3. Forecast assessment circular plots for each forecast horizon of 1-12 weeks ahead in the 30% of full forecast set. Individual sectors in each circular plot represent one disease category: their length represents the minimum error statistic value measured by mean absolute percentage error, while their color represents the corresponding forecast model type (naïve, simple, simple combinations, or combinations through supervised learning). Consequently, each individual sector denotes the type of forecast model that performs the best for the disease category at the particular forecast horizon. The axes of all circular plots are normalized to the 0 and maximum value (18.631%) of minimum error statistics across all disease categories and forecast horizons.

Having known that simple combinations outperform submodels for most disease causes and forecasting windows, we need to ascertain whether they generate statistically significantly superior forecasts. Thus, we conduct the DM test pairwise between forecasts generated by individual submodels and simple forecast combinations across the 16 disease admission categories. We record the proportion of statistically non-equivalent forecast performance for each disease category at each forecast horizon; a non-equivalent scenario would indicate that one of the simple combination schemes performed statistically better than a submodel in the particular forecast horizon and ED admission category. This procedure demonstrates that combining forecasts leads to statistically better point forecast performance versus simpler submodels for all forecast horizons (Table 1). The proportion of scenarios in which the simple combinations significantly outperform the submodel is more than half for all disease causes and forecast windows in our study setting. We also observe an increase in the proportion with increasing forecast horizon in some disease categories, for example, for admissions related to malignant neoplasms this proportion increases from 0.72 to 0.81 and for admissions related to neurological and sense disorders it goes up from 0.58 to 0.81 respectively for the 1- to 12-week ahead horizons.

| Disease category | Non-equivalence proportion by horizon | | | | | | | | | | | |
| --- | --- | --- | --- | --- | --- | --- | --- | --- | --- | --- | --- | --- |
| | 1 | 2 | 3 | 4 | 5 | 6 | 7 | 8 | 9 | 10 | 11 | 12 |
| Cardiovascular disease | 0.58 | 0.64 | 0.55 | 0.58 | 0.58 | 0.61 | 0.69 | 0.67 | 0.64 | 0.59 | 0.69 | 0.59 |
| Chronic respiratory disease | 0.73 | 0.63 | 0.72 | 0.72 | 0.75 | 0.75 | 0.73 | 0.73 | 0.70 | 0.69 | 0.73 | 0.72 |
| Factors influencing health status and contact with health services | 0.66 | 0.70 | 0.72 | 0.69 | 0.67 | 0.72 | 0.66 | 0.72 | 0.70 | 0.69 | 0.66 | 0.73 |
| Digestive disease | 0.77 | 0.73 | 0.73 | 0.77 | 0.75 | 0.75 | 0.75 | 0.75 | 0.77 | 0.75 | 0.73 | 0.73 |
| Endocrine disorders | 0.75 | 0.80 | 0.73 | 0.78 | 0.75 | 0.77 | 0.73 | 0.70 | 0.69 | 0.69 | 0.75 | 0.73 |
| Malignant neoplasms | 0.72 | 0.70 | 0.75 | 0.75 | 0.73 | 0.69 | 0.73 | 0.72 | 0.70 | 0.69 | 0.81 | 0.81 |
| Diabetes mellitus | 0.86 | 0.72 | 0.78 | 0.78 | 0.75 | 0.75 | 0.77 | 0.75 | 0.73 | 0.77 | 0.72 | 0.77 |
| Genitourinary disorders | 0.58 | 0.64 | 0.58 | 0.63 | 0.66 | 0.64 | 0.69 | 0.78 | 0.80 | 0.72 | 0.72 | 0.66 |
| Musculoskeletal disease | 0.64 | 0.64 | 0.63 | 0.64 | 0.67 | 0.70 | 0.67 | 0.69 | 0.72 | 0.73 | 0.67 | 0.64 |
| Infectious and Parasitic Diseases | 0.72 | 0.67 | 0.73 | 0.70 | 0.72 | 0.66 | 0.66 | 0.67 | 0.67 | 0.69 | 0.69 | 0.64 |
| Ill-defined diseases | 0.70 | 0.66 | 0.63 | 0.70 | 0.70 | 0.72 | 0.73 | 0.78 | 0.84 | 0.83 | 0.77 | 0.77 |
| Neurological and sense disorders | 0.58 | 0.59 | 0.56 | 0.63 | 0.64 | 0.61 | 0.66 | 0.66 | 0.70 | 0.81 | 0.75 | 0.69 |
| Oral Diseases | 0.55 | 0.64 | 0.63 | 0.61 | 0.63 | 0.66 | 0.67 | 0.67 | 0.61 | 0.63 | 0.64 | 0.61 |
| Other neoplasms | 0.53 | 0.58 | 0.56 | 0.64 | 0.53 | 0.56 | 0.63 | 0.61 | 0.67 | 0.66 | 0.58 | 0.58 |
| Respiratory Infection | 0.78 | 0.84 | 0.77 | 0.78 | 0.81 | 0.80 | 0.78 | 0.78 | 0.78 | 0.77 | 0.75 | 0.75 |
| Skin diseases | 0.61 | 0.63 | 0.64 | 0.66 | 0.66 | 0.59 | 0.64 | 0.59 | 0.63 | 0.59 | 0.59 | 0.59 |

**Table 1** The proportion of non-equivalence scenarios (simple forecast combinations are superior to simple submodels) obtained using the Diebold-Mariano (DM) test statistic out of a total of 64 comparison scenarios for all 16 simple submodels (including the naïve model) with 4 simple combinations. The DM test is used to test the statistical equivalence of the forecast errors for the different models and the different forecasting horizons for each disease category. This is computed for the forecast residuals in the full forecast set across the 12-week forecast horizon.

There are also very significant differences in the forecast performance of different forecast combinations. Simple forecast combinations, especially Bates and Granger (**P3**) consistently outperform the other combinations based on supervised learning across all disease causes except infectious and parasitic diseases for the 1-week ahead (Fig. 4, MAPE: 4.287 – 18.631%), 6-weeks ahead (Fig. 4, MAPE: 5.244 – 17.611%), and 12-weeks ahead forecast horizons (Fig. 4, MAPE: 5.035 – 15.476%). Notably, simple forecast combinations have a lower tendency to mean-revert, resulting in comparatively lower error rates at longer forecast horizons (See S2 file). The Bates and Granger (**P4**) scheme performs slightly better than both equal weights scheme and median forecast scheme for the 1-week ahead (Fig. 4, MAPE: 5.155 – 20.825%), 6-weeks ahead (Fig. 4 here, MAPE: 6.641 – 20.150%), and 12-weeks ahead forecast horizons (Fig. 4, MAPE: 6.214 – 18.460%), as the expanding window may be able to adjust weights based on submodel's past performance and prioritize better-performing submodels in the forecast combinations.

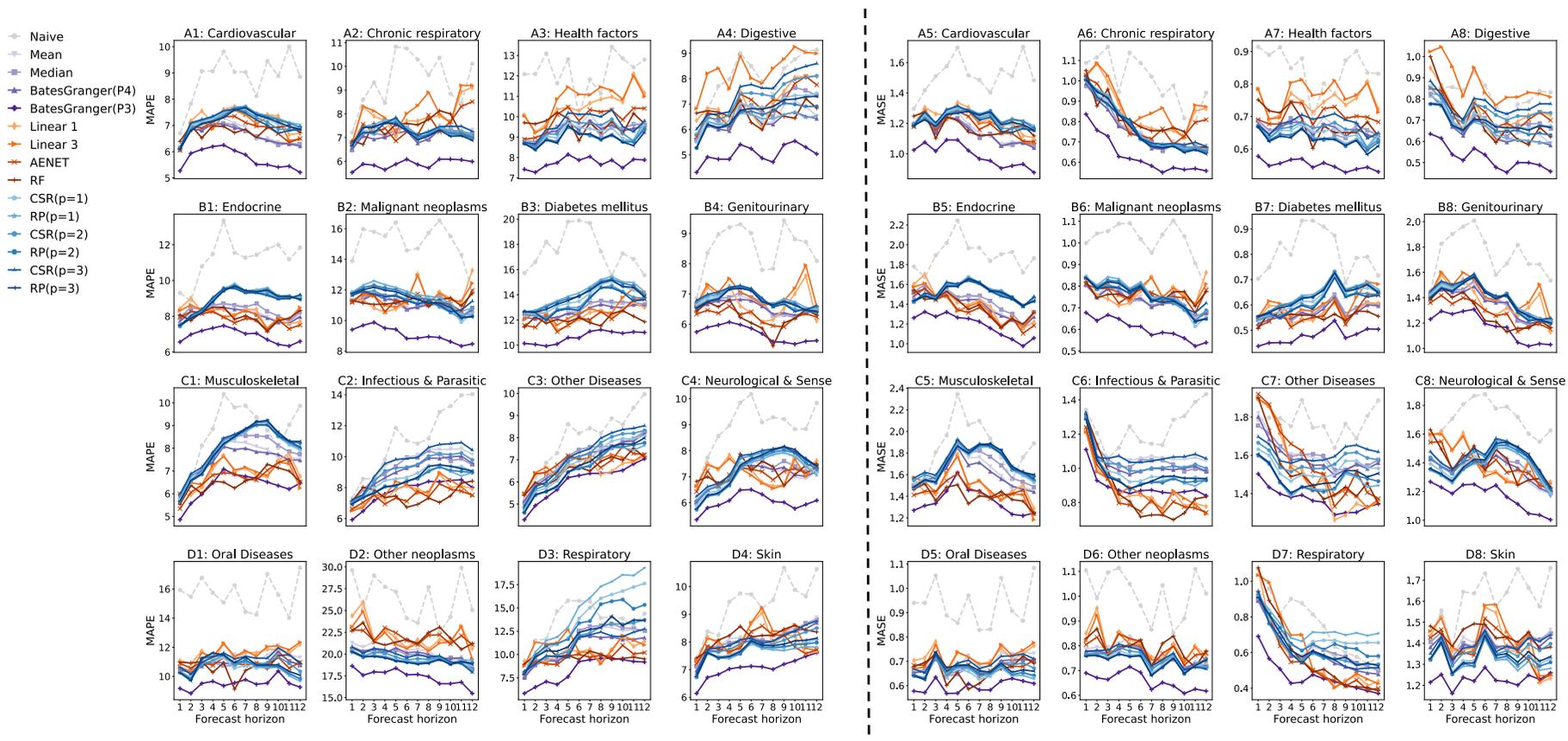

Fig. 4. Forecast assessment statistics in the 30% full forecast period. Mean absolute percentage forecast error (MAPE) (left of the dotted line) and mean absolute scaled error (MASE) rates (right of the dotted line) of the 15 forecasting models which include 1 naïve model, 4 simple point forecast combination models: equal weights (Mean), median forecast (Median), weights based on an expanding window of most recent data (BatesGranger (P4)) and weights based on contemporaneous submodel's forecast errors (BatesGranger (P3)), across the forecast horizon of 1-12 weeks ahead for individual disease categories, and 6 forecast combinations through high-dimensional supervised learning: 2nd-order linear regression (Linear 1), 2nd-order constrained linear regression (Linear 3), 2nd-order adaptive elastic net (AENET), 2nd-order random forest (RF), complete subset regressions (CSR) and random projections (RP), across the forecast horizon of 1-12 weeks ahead for individual disease categories, where there are 3 cases for CSR and RP respectively ($p = 1, 2, 3$). Because the forecasting performance of 2nd-order linear regression omitting intercept term (Linear 2) is much lower than all other schemes, it is not included in this visualization.

Our work further extends the evaluation of combination forecasts to include an assessment of supervised learning combinations schemes. However, as we generate supervised learning forecast combination weights on actual observations and the forecast outputs from individual submodels, we generate a smaller set of forecast outputs and calculate the validation period on this smaller time window (30% of the size of forecast dataset for individual submodels) across all submodels and combination schemes. In general, for most ED causes, the predictive accuracy of the supervised learning forecast combinations is similar to that of the other simple forecast combinations except for Bates and Granger (**P3**). In particular, for infectious and parasitic disease and respiratory infections, we note that supervised learning forecast combinations significantly outperform the other three simple forecast combinations and even surpass Bates and Granger (**P3**) in some forecasting windows. Among the linear regression combinations, the regression with coefficients constrained to sum to 1 (Linear 3) has a slightly higher prediction error than the unconstrained linear regression (Linear 1), but not a particularly large difference. The adaptive elastic net (AENET) outperforms the above linear schemes without regularization penalties for almost all causes and all horizons. The nonlinear supervised learning combinations, on the other hand, outperform the linear schemes in most disease categories, and improvements are more apparent as the forecast horizons increase.

The forecast performance of complete subset regressions (CSR) and random projections (RP) is generally comparable to that of equal weights and median forecast schemes in simple combinations. In some disease causes such as chronic respiratory disease, factors influencing health status and contact with health services, digestive disease, and other neoplasms, the forecasting performance is better than several supervised learning forecast combinations (Fig. 4A2, 4A3, 4A4, 4D2). Moreover, we do not observe significant differences between different $p$ for the same scheme. However, for the same $p$, RP is always slightly better than CSR.

To explore how the forecast combinations integrate the strengths of the different submodels, we visualize how the weights of the different submodels in the forecast dataset change over time in different forecast horizons. Fig. 5 illustrates the Bates and Granger (**P4**) for the four diseases as well as for the 1,4,8 and 12-week ahead forecast horizons (see S1 text for all categories and **P3**). We observe that the weights of autoregression with exogenous variables were consistently low, which is consistent with poor model forecast performance (Fig. 2, 5). The weights of naïve forecasts decrease with increased epidemiologic weeks for the 8-week and 12-week ahead forecast horizons, concomitant with its good performance in short forecast horizons and degrading predictive ability in the long-term horizons (Fig. 2, 5). The weights of the nonparametric machine learning models XGBoost, lightGBM, KNN, and RF are high across all forecast horizons and all four categories. The contributions of elastic net and adaptive elastic net are also high for forecast combinations for ED admissions related to cardiovascular and musculoskeletal disease. However, the latter does not contribute as much to the forecast combinations as the former (Fig. 5).

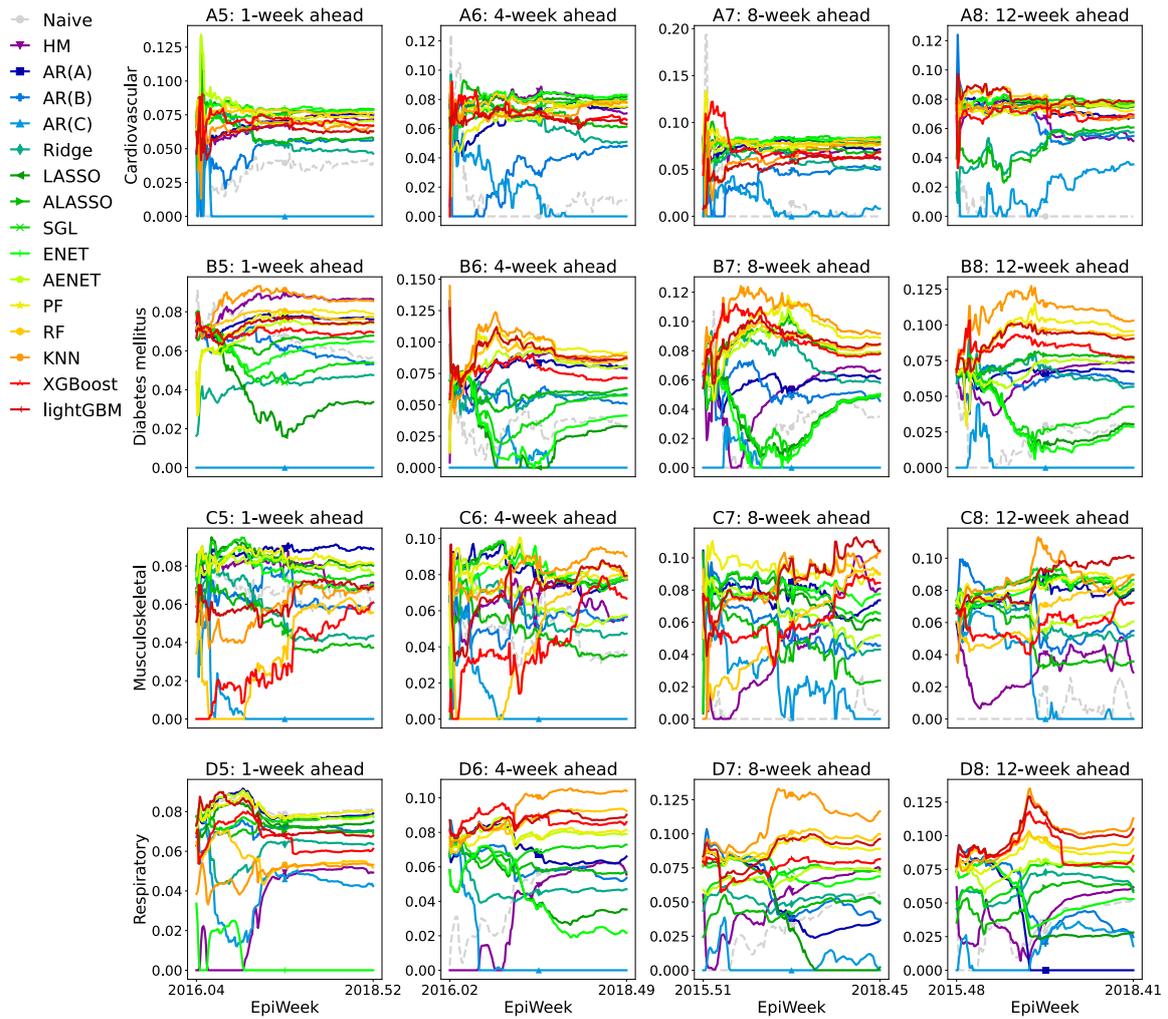

Fig. 5. Visualization of how the weights of individual submodels in the Bates and Granger (**P4**) change over epidemiological weeks for 4 disease categories: Cardiovascular disease, Diabetes mellitus, Musculoskeletal disease, Respiratory Infection and 4 forecasting windows (1-week, 4-week, 8-week, 12-week).

## Discussion and conclusions

In contrast to prior work, our goal is to develop short to long term forecasting models for ED admissions by their admission causes, which can help resource planners on forward guidance of heightened demand loads by specific disease subgroups. We demonstrate that incorporating many relevant predictors into forecasting models does not significantly degrade performance and can even enhance forecast accuracy for some models. Moreover, cause-specific forecasts provide more granular insights into ED demand by cause and induces higher forecast accuracy compared to all-cause forecasting (Fig. 1). Therefore, forecasting the total number of ED admissions may not be able to provide the requisite resolution and lead to a suboptimal distribution of medical resources, wasted resources and delayed treatment times for patients[26]. Here, we test multiple forecasting techniques based on nationwide data on weekly cause-specific emergency department admissions and a large amount of environmental and anthropogenic data, and develop multiple forecast combination schemes based on submodel forecasts. And we evaluate the absolute and relative performance of the models based on two error metrics, as well as statistically characterizing the equivalence of the combined and submodel forecasts. Finally, we perform forecasting exercises to evaluate how including exogenous factors and aggregating cause-specific ED forecasts affect the accuracy of total ED admissions predictions. We demonstrate that the proposed forecast combinations show relatively stable and good point forecast performances even as the forecast horizon increases in the study period (Fig. 4). Also, incorporating relevant predictors and forecasting for different causes enhances performance and offers finer insights for decision-making. Therefore, we develop forecast combination schemes that can aid in ED resource allocation and planning over long horizons with low error rates.

Specifically, forecast combinations demonstrate large, statistically significant improvements in point forecasting upon individual submodels (Fig. 3, 4, Table 1). This pattern is shown across most ED admission categories and forecast horizons. By constructing forecast combinations which adaptively combined many constituent submodels based on their most current predictive performance in out-of-sample data (Fig. 4), we generate combined forecasts that have improved point forecast accuracy over constituent submodels, despite some submodels demonstrating degrading performance over longer forecast horizons. However, heterogeneities in the trajectories of ED admissions and the predictors for these trajectories are apparent when compared across each admissions category. The same forecasting model, when employed to different admission categories, would consequentially have varying performance when subject to forecast assessments (Figure 2–4). In example, AR(B) and AR(C), which perform poorly in other ED admission categories, outperform AR(A) for ED admissions due to respiratory infections (Fig. 2), which demonstrates that the direct inclusion of all exogenous covariates may lead to better forecasts in some cases.

Comparison between submodels and forecast combinations also reveals several trends. Forecast combinations do not mean-revert as quickly as submodel constituents (See S2 file). Combining forecasts also yields a relatively low error rate, especially over longer forecast horizons when compared versus submodel constituents, which can be attributed to the simple forecast combinations aggregating forecasts across all 16 submodels and minimizing extreme errors in poorly performing submodels. This is concordant with past literature which shows that large predictive performance improvements can be yielded from combining forecasts[27]. Simple forecast combinations which are constructed using weighting schemes may already possess some of these properties, by removing misspecification bias and instability generated from forecasting submodels as well as reducing the Mean Squared Forecast Error (MSE) ratios associated to these forecasting submodels[28]. Similarly, the AENET (**P8**) in the supervised learning scheme adaptively drops poor performers. As the performance of the equal weights scheme (**P1**) has been shown to depend heavily on prior model instability and difficulty in identifying the optimal model weights[29], other weights derived from more complicated forecast combination schemes can still provide improvements over submodel constituents if weights can be accurately identified (Fig. 3, 4). Bates and Granger (**P3**) performs better than Bates and Granger (**P4**) as weights are calculated directly from the current data rather than being based on all past and contemporaneous out-of-sample data (Fig. 4) This may be due to the former method being able to avoid the instability of the prior submodels and that the computation of its combination weights is not as dependent on past forecast errors as the latter.

As we use cross-validation and out-of-sample information containing only additional one epidemiological week at each iteration, it is difficult to adequately estimate the covariance matrix, and "optimal" estimation errors tended to be particularly large, especially when there were many forecasts to be combined. Therefore, for Bates and Granger (**P3**) and (**P4**), we compute the optimal weights under the MSE loss[30]. However, the properties of the optimal forecasts obtained through the MSE loss are usually not robust under more general assumptions regarding the loss function[31]. Future work can explore more robust loss functions under the same experimental setting as an improvement.

Further exploration of more complicated forecast combinations through supervised learning suggests that these methods may also yield large improvements versus individual simple submodels. This is done by combining forecasts based on past errors using either linear regression, or ensemble machine learning methods. As adaptive elastic net and random forest outperform other supervised learning forecast combinations, there may be a nonlinear relationship between the submodels' forecasts and actual observations. In addition, in the other two combination schemes based on supervised learning followed by averaging, RP consistently slightly outperforms CSR for the same $p$. The possible reason is that when the same $p$ is applied to both, the dimensionality of the data after dimensionality reduction will not be the same. This is because CSR selects the different subsets of exogenous predictors that contain all 8 lag orders, while RP performs dimensionality reduction among all exogenous predictors. For example, when p=1, CSR selects 8 predictors in each of the environmental variables and other disease ED variables, while RP projects the high-dimensional environmental data and other disease ED data into 1 new predictor. So RP results in fewer predictors in the final regression model, which may avoid the overfitting that may result from inclusion of multiple independent variables. In addition, RP compresses the exogenous information into a low-dimensional space via a random projection matrix. In each projection, each exogenous variable loses part of its information randomly and consequentially preserves the majority of the original data in the random projection. Whereas CSR removes some variables that may be significantly associated with the forecasting target in each subset.

There are several limitations in this study. (**a**) We only collect data through 2018, a period in which there are no structural breaks in transmission for all disease categories. As the COVID-19 pandemic may have caused a drastic change in ED demand[32], we could not assess whether our methods are robust over periods where structural breaks exist. Future work should assess the applicability of these forecasting tools in longer time periods and over other study sites – to ensure both temporal and geographical generalizability. In addition, we only present the prototypes developed based on retrospective data. Further embedding these prototypes into the hospital's infrastructure and linking them to the hospital's database and weather API allow model parameters to be updated based on the latest data. For example, the hospital can retrain the model every month or quarter to ensure that hospital managers always use the latest predictive model for real-time admission forecasting. (**b**) Although many predictors are included in this study, unobserved confounding may bias our forecasts. Therefore, it may be necessary to perform an epidemiological analysis before conducting forecasting to infer variables that may be associated to each cause-specific ED admission category and to reduce forecast noise. (**c**) Our results demonstrate that Bates and Granger (**P4**) does not outperform the equal weights (**P1**) scheme, mirroring the "forecast combination puzzle" phenomena. Due to the randomness of the combination weights, it is difficult to make statistical inferences about weighted combined forecasts, especially from their sampling distributions to obtain standard errors. Therefore, future work may explore statistical inference for forecast combinations. (**d**) In this study, (**e**) We have only considered point forecast combinations and not density forecast combinations. However, the uncertainty of the forecasts is difficult to ignore, and it can give decision-makers an idea of how much confidence we have in the forecasts. Future work should explore the utility of density forecast combinations in characterising uncertainty in forecasting cause-specific ED admissions.

## Methods

**Data** The average daily ED admissions case counts are collected by hospitals across Singapore. We categorize admissions according to the Singapore Burden of Disease categories[33]. ED admissions data is recorded daily and aggregated by disease category weekly to the Epidemiological Week (EW) timescale from EW 1 of 2009 to EW 52 of 2018. We remove disease categories where case counts exhibit case classification errors or inconsistencies over time (sudden consistent zero counts or change in classification categories) from the analysis (See S1 text for full list of data and data sources). Therefore, we only include 16 disease categories in this study setting and ED admissions of all causes seem to vary with no clear pattern.

Given the well-studied effects of the natural environment on many of our disease categories, we condition our ED forecasts on a detailed list of meteorological and environmental variables. Meteorological data is obtained from ERA5-Land, published by the European Centre for Medium-Range Weather Forecasts. ERA5-Land provides hourly estimates across a 30-kilometer grid, which we aggregate to the epidemiological week timescale and spatially average across the land area of Singapore. We calculate maximum, mean, and minimum air temperatures at 2 meters (Kelvin) to represent thermal forcing and stress on host populations, while total weekly rainfall (meters) serves as a proxy for its influence on population mixing behavior and time spent outdoors. We utilize air temperature and dew point temperatures to calculate relative humidity (%) and absolute humidity (g/m3) using standard formula[34]. Ambient air pollutant data is obtained from Air Quality Open Data Platform (AQICN), which provide daily timescale data reported at 5 surface level sites across Singapore. The daily mean pollutant data for Singapore is calculated by averaging the data from the 5 sites, then aggregated to the epidemiological week timescale. Particulate matter indicators (PMI) PM2.5 and PM10 indicate the level of fine inhalable particles and inhalable particles respectively, while pollutant concentrations of $O_3$, $NO_2$, $SO_2$ and CO indicate the amount of hazardous air pollution present in the urban environment. We convert the values from AQI index levels to their respective concentration units using the breakpoints and formulas provided by the United States Environmental Protection Agency[35]. All environmental variables are also relatively constant. Only PM2.5 and PM10 exhibit significant peaks due to seasonal haze pollution with average values of 18.367μg/m³ (Range: 9.164μg/m³ — 167.028μg/m³) and 29.965μg/m³ (Range: 16.664μg/m³ — 213.948μg/m³) for PM2.5 and PM10 respectively.

**Individual forecast models** We generate point forecasts for ED admissions using 16 different submodels. In all but 4 models, we condition our forecast on a set of exogenous predictors that include 12 environmental variables along with the ED admission case counts of the other disease categories (15 diseases). We characterize the temporal dynamics of the dependent variable of interest by including 8 weeks of lags for all the predictor variables including that of the disease of interest, which results in a high-dimensional set of 224 predictors. Since near to medium-term forecasts of ED admissions is crucial for real-time resource optimization in hospitals, we forecast ED admissions demand load for the 1- to 12-week ahead horizon for all 16 disease categories using a rolling window.

Here, we detail the individual submodels. Let the ED admissions case counts of disease $D$ at time $t+h$ be $y_{t+h,D}$, where $h$ is the forecast horizon. We first consider 2 parsimonious forecast models that do not rely on any exogenous predictor information.

*Random walk*

Suppose disease case counts are unpredictable and follow a random walk:

$$y_{t+1,D} = y_{t,D} + \varepsilon_{t+1,D} \tag{1}$$

and $\varepsilon_{t+1,D}$ is a martingale difference series disturbance (mds) term. The naïve forecast of $y_{t+1,D}$ is therefore $E[y_{t+1,D}|I_t] = y_{t,D}$. Furthermore, it is straightforward to see that this holds for any forecast horizon: $E[y_{t+h,D}|I_t] = y_{t,D}$, which in other words mean that the model-implied forecast of future case counts is simply its presently observed value. Due to the simplicity of the naïve forecast and the implicit null hypothesis of unpredictability, we shall treat this as our baseline or benchmark forecast.

*Historical mean*

Another parsimonious model is that of the historical mean or average. We construct the forecast using the rolling average: $E[y_{t+h,D}|I_t] = \frac{1}{J}\sum_{j=0}^{J-1} y_{t-j,D}$ where the window of size $J$ was prespecified to be lag order, i.e. 8.

*Autoregressive model*

We consider an autoregression with and without exogenous variables which involves a standard linear projection for the $h$-week ahead forecast of ED admissions demand load for each disease category. The forecasting projection for ED admissions due to disease category $D$ at $t+h$, including 8-week lags of exogenous predictors, is given as:

$$y_{t+h,D} = \beta_0 + \sum_{l=0}^{7} \beta_{D,l} y_{t-l,D} + \sum_{k=1}^{K}\sum_{l=0}^{7} \beta_{k,l} x_{t-l,k} + \sum_{d=1, d\neq D}^{16}\sum_{l=0}^{7} \beta_{d,l} y_{t-l,d} + u_{t+h,D} \tag{2}$$

where autoregressive terms include past and contemporaneous observations of cause-specific ED admissions $y_{t-l,D}$, environmental forcing on future ED admissions burden is incorporated through $K$ environmental covariates $x_{t-l,k}$. Other ED admissions for different causes are denoted as $y_{t-l,d}$ and the forecast error term is given by $u_{t+h,D}$. $\beta_{D,l}, \beta_{k,l}$ and $\beta_{d,l}$ are the set of regression coefficients. Under mild conditions on the autoregressive process of $y_{t,D}$, the $h$-week ahead point forecast is given by the conditional mean. We estimate 3 specifications on exogenous variables (**A**) without any exogenous variables, (**B**) with only lagged environmental variables and (**C**) with both lagged environmental variables and ED admissions for other different disease categories. We abbreviate these three sub-models as AR (A), AR (B), AR (C). AR (C) is motivated by potential spillover effects between ED admissions disease categories due to the finite supply of ED resources.

A potential key problem of directly estimating (2) via least squares regression is the large number of parameters involved, which can entail large estimation uncertainty. To address this, we consider 5 regularization strategies : (i) ridge regression, (ii) least absolute shrinkage and selection operator (LASSO), (iii) elastic net, (iv) adaptive LASSO, and the (v) sparse group LASSO.

*Ridge regression*

We first represent our data in matrix-vector notation. Define $\mathbf{y_D} = [y_{9+h,D}, \ldots, y_{T+h,D}]'$, $\boldsymbol{\beta_D} = [\beta_0, \beta_{D,0}, \ldots \beta_{D,7}, \beta_{k,0}, \ldots, \beta_{K,7}, \beta_{d,0}, \ldots, \beta_{d,7}]'$, and $\mathbf{X_D}$ is the data matrix of the corresponding predictors in (2). The ridge regression estimator solves the following penalized least squares problem:

$$\underset{\boldsymbol{\beta_D}}{argmin} |||\mathbf{y_D} - \mathbf{X'_D}\boldsymbol{\beta_D}|||_2^2 + \lambda_D \|\boldsymbol{\beta_D}\|_2^2 \qquad (3)$$

where $\|.\|_2$ is the $\ell_2$ norm, and $\lambda_D$ is the ridge shrinkage parameter (for disease *D*). We obtain the obtain the optimal penalty term for the ridge estimator (and all the regularization techniques below) using a time series cross validation scheme (5 iterations) instead of splitting samples in random folds as training and validation sets to respect the time-dependent nature of our data[36].

*LASSO*

The LASSO estimator solves the following problem:

$$\underset{\boldsymbol{\beta_D}}{argmin} |||\mathbf{y_D} - \mathbf{X'_D}\boldsymbol{\beta_D}|||_2^2 + \lambda_D \|\boldsymbol{\beta_D}\|_1 \qquad (4)$$

where the $\ell_1$ norm in the penalty term is given as the sum of the absolute values of all the elements in $\boldsymbol{\beta_D}$. The $\ell_1$ penalty, relative to the ridge penalty, can induce sparsity by setting some of the coefficients to be equivalent to 0. Intuitively, this means that predictor might not be relevant in forecasting the target variable. The degree of sparsity is determined by the value of the penalty term (i.e. the larger the term the sparser the solution is), and again we obtain this using a data-driven cross-validation scheme.

*Adaptive LASSO (ALASSO)*

The vanilla lasso suffers from several drawbacks. If a group of predictors is highly correlated with one another, the LASSO may pick only one predictor out of this group (i.e. assign it a non-zero coefficient) and set the rest of the coefficients to 0. Furthermore, on a theoretical level, to guarantee that the LASSO accurately selects the relevant predictors, we require the relevant and irrelevant predictors to be orthogonal. This is unreasonable in our setting given that many of the environmental and disease-related data are highly correlated. One major factor contributing to these drawbacks is the fact that we are imposing a uniform penalty on all of the predictors in the LASSO problem. To relax this, we can consider assigning individual penalties to each regressor, which leads us to the adaptive LASSO[37]. The adaptive LASSO solves the following problem:

$$\underset{\boldsymbol{\beta_D}}{argmin} |||\mathbf{y_D} - \mathbf{X'_D}\boldsymbol{\beta_D}|||_2^2 + \lambda_D \sum_{i=1}^{P} w_{D,i} |\beta_{D,i}| \qquad (5)$$

where $\beta_{D,i}$ is the i-th element in the $P \times 1$ vector $\boldsymbol{\beta_D}$ and $w_{D,i}$ is its associated weight which is specified prior to estimation. We use the following weight specification:

$$\widehat{w}_{D,i} = \frac{1}{|\hat{\beta}_{D,i}|} \qquad (6)$$

where $\hat{\beta}_{D,i}$ is the corresponding estimate from submodel AR (C) which is the unpenalized least squares regression estimate. The logic of this is that if a predictor is not useful in forecasting the target variable in the unrestricted autoregressive model, then its coefficient $\hat{\beta}_{D,i}$ should be close to 0, and thus the assigned weight would be very large. Hence, the corresponding penalty would be large, and the adaptive LASSO would more likely drop the predictor.

*Sparse group LASSO (SGL)*

Group LASSO allows sparsity to be induced by a predefined group of covariates. On top of the between-group selection, we use sparse group LASSO to induce within-group sparsity, in order to select the most meaningful predictors from the most meaningful groups[38]. Here, we implement groups by all types of exogenous variables $1, \ldots, G$, where each group is defined to be a single variable and all its lags, so that variables in each variable group can be jointly included or removed from the regression. The sparse group LASSO solves:

$$\underset{\boldsymbol{\beta_D}}{argmin} \|\boldsymbol{y_D} - \boldsymbol{X'_D}\boldsymbol{\beta_D}\|_2^2 + \alpha\lambda_D\|\boldsymbol{\beta_D}\|_1 + (1-\alpha)\lambda_D \sum_{g=1}^{G} \sqrt{d_g}\|\boldsymbol{\beta_{D,g}}\|_2 \tag{7}$$

where $\boldsymbol{\alpha\lambda_{D1}}$ is the parameter-wise penalty, and $(\mathbf{1}-\boldsymbol{\alpha})\boldsymbol{\lambda_{D2}}$ is the penalty for groups. $\boldsymbol{\alpha} \in [\mathbf{0},\mathbf{1}]$. It is a convex combination of the LASSO and group LASSO penalties. $\boldsymbol{\beta_{D,g}}$ is the vector of coefficients corresponding to the $g$-th group, $d_g$ is the length of the vector, and $G$ is the total number of predictor groups.

*Elastic net (ENET)*

The elastic net overcomes limitations related to ridge regression and LASSO respectively, such as high model complexity despite parameter shrinkage for high-dimensional ridge regression model, and poor variable selection among a group of highly correlated variables for LASSO. The latter which will likely occur when selecting exogeneous predictors in determining ED admissions demand load, due to their tendency to co-move. Elastic net combines both penalty terms in ridge and LASSO regression, resulting in the following objective function:

$$\underset{\boldsymbol{\beta_D}}{argmin} \left\|\boldsymbol{y_D} - \boldsymbol{X'_D}\boldsymbol{\beta_D}\right\|_2^2 + \lambda_{D,R}\|\boldsymbol{\beta_D}\|_2^2 + \lambda_{D,L}\|\boldsymbol{\beta_D}\|_1 \tag{8}$$

where $\lambda_{D,R} = (1-\alpha)\lambda_D$, $\lambda_{D,L} = \alpha\lambda_D$, and $\alpha \in [0,1]$. Clearly, the elastic net penalty is a convex combination of the ridge and LASSO penalties.

*Adaptive elastic net (AENET)*

The adaptive elastic net replaces the original LASSO penalty with the adaptive LASSO penalty and combines it with the regular ridge penalty to enable the elastic net to exhibit oracle properties[39].

*Factor model (PF)*

As an alternative to predictor selection. We can posit that our large number of predictors can be explained by a small number of common factors. Instead of using all of the predictors, we extract the small number of common components of all exogenous variables, and use them together with all the lags of the target cause' ED admissions to forecast our target variable. This is the methodology proposed by[40]. To formalize this, let $g_{t,D} = (x_{t,1}, \ldots, x_{t-7,1}, \ldots, x_{t-1,K}, y_{t,1}, \ldots, y_{t-7,16})'$ be a collection of the exogenous variables and the cross-disease counts. The factor model can be written as:

$$g_{i,t,D} = \lambda_{i,D}'f_{t,D} + e_{i,t,D} \tag{9}$$

$$y_{t+h,D} = y'_{t,D}\beta + f'_{t,D}\gamma + z_{t+h,D} \tag{10}$$

where $g_{i,t,D}$ is the $i$-th element of $g_{t,D}$, $f_{t,D}$ is a $R \times 1$ vector of $R$ common but latent factors or components, and $\lambda_{i,D}$ is the corresponding vector of how the $i$-th predictor reacts to the common components, also known as the factor loading. The second equation (Eq. (10)) is the forecasting equation, and it is a projection of $y_{t+h,D}$ on the common components, with $\gamma$ being the projection coefficients and $\beta$ being the coefficients of the autoregression terms of the target disease as represented by $y_{t,D}$.

As the common factors are latent, we estimate them from the cross-section of predictors using principal components analysis (PCA) which yields a consistent estimator of the underlying factor space. We take the first several principal components which could explain at least 85% of the variances. The second equation is estimated via least squares by replacing the latent factors with the estimated PCA components.

*Random forest (RF)*

In addition to linear models, we consider nonparametric models to be estimated with ensemble methods such as random forest, gradient boosting machines and k-nearest neighbours. Here, our forecasting equation is given as

$$y_{t+h,D} = f(g_{t,D}, y_{t,D}) + \varepsilon_{t+h,D} \equiv f(q_{t,D}) + \varepsilon_{t+h,D}, \tag{11}$$

where $f(.)$ is an unknown and potentially non-linear function.

We first begin with a discussion on regression trees. For simplicity, assume that we only have two predictors for disease category $D$ given as $x_{1,t,D}$ and $x_{2,t,D}$. Situate the predictors $(x_{1,t,D}, x_{2,t,D})$ in the x-y plane of a 3-dimensional coordinate plane, with the response variable $y_{t+h,D}$ on the z-axis. The regression tree will find the optimal partition of the $(x_{1,t,D}, x_{2,t,D})$ plane such that the dependent variable is constant within each partition. This can be achieved via some recursive binary partitioning algorithm. The estimator of the unknown function via a regression tree is given as:

$$\hat{f}(q_{t,D}) = \sum_{j=1}^{J} \hat{c}_j \, \hat{1}_j\{q_{t,D}\} \tag{12}$$

where $\hat{c}_j$ is the sample average of $y_{t+h,D}$ within partition $j$, $R_j$, and $\hat{1}_j\{q_{t,D}\} = 1$ if $q_{t,D} \in R_j$ and 0 otherwise.

The random forest would then be an average of many regression trees trained on bootstrap samples. If we let $M$ to be the total number of bootstrap iterations, the random forest estimator can be written as:

$$\hat{f}(\boldsymbol{q}_{t,D}) = \frac{1}{M} \sum_{m=1}^{M} \left( \sum_{j=1}^{J_m} \hat{c}_{j,m} \, \hat{1}_{j,m}\{q_{t,D}\} \right), \tag{13}$$

where objects with the subscript $m$ are obtained from a specific bootstrap iteration. We independently fit 1000 regression trees to resampled subsamples and we assess the quality of regression tree splits using the mean squared error criterion. We set the maximum number of predictors in the partitioning consideration to be the square root of the total number of predictors. We aggregate the individual regression tree fits using a simple average as detailed above to control for over-fitting.

*K-nearest neighbours (KNN)*

KNN is a nonparametric method that generates forecasts by assigning the prediction to be the average of the values of the nearest neighbours (in a sense to be elaborated below). $k$ is the parameter that determines how many nearest neighbours to consider. A larger $k$ will result in a smoother model, which reduces variance but increasing bias and vice-versa. We execute the KNN algorithm as follows:

1. Suppose we are at time $T$ and want to make a h-step ahead forecast of $y_{T+h,D}$ using our predictors $q_{T,D}$.
2. We compare the distance, as measured by the $\ell_2$ norm, between $q_{T,D}$ and $q_{s,D}$ for all $s < T$. Let $t_1, \ldots, t_k$ represent the time indices that yield the $k$ smallest distances.
3. Then, our forecast of $y_{T+h,D}$ is given by the following average: $\frac{1}{k}\sum_{j=1}^{k} y_{t_j+h,D}$.

We repeat this process for all horizons and disease categories. In our study, we set $k = 5$.

*Gradient Boosting Machines (GBM)*

GBM is similar to random forest, but the regression trees are iteratively fitted to the data, which means the trees are built successively. All boosting techniques are characterized by an additive training process, where a new weak learner is incorporated into the model at each iteration. Specifically, in GBM, the weak learner introduced at each step is a decision tree. This boosting process is shown as follow:

$$f_n(q_{t,D}) = f_{n-1}(q_{t,D}) + h(q_{t,D}), \tag{14}$$

where $f_n(.)$ is the latest model after $n-1$ rounds and $h(.)$ is the new decision tree that is added into the model. We initialize with fitting a singular decision tree. Subsequent regression trees ($n=1000$) are then fitted to the discrepancy between the predictions generated by the previous tree and actual observations. This is continued until the discrepancy between predictions and data crosses a pre-specified error threshold. Two specific algorithms are separately used (XGBoost[41], LightGBM[42]) to generate forecasts.

**Forecast combinations** As discussed earlier, a key issue in forecasting ED admissions is that we do not know a priori which forecasting model works best, for which disease, and whether its performance would be stable over time. To hedge against significant forecast model uncertainty, we consider several forecast combination schemes here. The main idea is simple, we first generate forecasts from the 16 individual sub-models, and instead of singling out a single forecast, we aggregate the full range of forecasts to obtain a single point forecast. The aggregation technique differs depending on which forecast combination we employ. More formally, we consider the following parsimonious linear combination to produce a combined forecast:

$$w_1 \hat{y}_{T+h,D,1} + \cdots + w_{16} \hat{y}_{T+h,D,16} \equiv \boldsymbol{w}' \hat{\boldsymbol{y}}_{T+h,D} \tag{15}$$

where $\hat{y}_{T+h,D,i}$ is the forecast implied from the $i$-th submodel at time $T$, and $\boldsymbol{w}$ is a vector of forecast combination weights. These weights will vary according to the chosen combination method. Although not explicitly indicated here, for many of our approaches, the weights are adaptive i.e. $\boldsymbol{w} = \boldsymbol{w}_{T+h}$ (specific to the horizon), and will therefore vary over time. We do this to accommodate potential time-varying forecast performance of the individual submodels. Furthermore, note that we have suppressed the dependence of the weights on the target disease category $D$ for notational convenience.

*Equal weights (P1)*

We first generate simple combinations of the individual forecasts where each forecasting model is assigned equal weight: $\boldsymbol{w} = (\frac{1}{16}, \frac{1}{16}, \ldots \frac{1}{16})'$. The equal-weight combination is simple yet often very challenging to beat.

*Median forecast (P2)*

This simple approach assigns a weight of 1 to the median forecast and 0 elsewhere. Note that the median forecast can vary over time.

*Bates and Granger combinations*

We follow the popular approach proposed by Bates and Granger[43] to allocate a larger weight to forecasts that have small forecast errors and smaller weight to forecasts that are more inaccurate. Correlations between forecast errors are ignored by this scheme. Let the forecast errors from submodel $i$ be $\hat{e}_{i,t+h}$:

$$\hat{e}_{i,t+h} = \{\hat{y}_{t+h,D,i} - y_{t+h,D}\}^2, \tag{16}$$

we use a rolling window of the most recent $v$ observations and forecasts to compute the weights. The weight on the forecast from submodel $i$ is given as:

$$w_{i,T+h} = \frac{\left(\sum_{t=T-v+1}^{T} \hat{e}_{i,t+h}\right)^{-1}}{\left(\sum_{t=T-v+1}^{T} \hat{e}_{1,t+h}\right)^{-1} + \cdots + \left(\sum_{t=T-v+1}^{T} \hat{e}_{i,t+h}\right)^{-1} + \cdots + \left(\sum_{=t-v+1}^{T} \hat{e}_{16,t+h}\right)^{-1}}, \tag{17}$$

then, we consider 2 situations:

(1) **P3:** when $v = 1$, we calculate the relative performance weights at time $T$
(2) **P4:** when $v = T$, we use an expanding window involving the information only in the forecast set up to time $T$.

Subsequently, we consider more sophisticated approaches via a (un)constrained linear regression framework to learn the forecast combination weights[44].

*Linear Regression*

Suppose we wish to estimate $\boldsymbol{w}_{t+h}$, we use as response $y_{t+h,D}$ and the forecasts from the individual submodels $\{\hat{y}_{t+h,D,1}, \dots, \hat{y}_{t+h,D,16}\}$ within a regression framework. Here, we use 70% of the data from the forecast set for the fitting and the latter 30% for the evaluation of the performance of the forecast combinations (**P5–P9**). Therefore, $\boldsymbol{w}_{t+h}$ in (**P5–P9**) is not time-varying. Following the linear regression-based combinations proposed by Granger and Ramanathan[45], we consider three variations:

(1) **P5:** a linear regression of past outcomes on the forecasted values generated with no constraints, $y_{t+h,D} = \alpha + \boldsymbol{w}'_{t+h} \hat{\boldsymbol{y}}_{t+h,D} + \tilde{e}_{t+h}$;
(2) **P6:** omitting the intercept term, $y_{t+h,D} = \boldsymbol{w}'_{t+h} \hat{\boldsymbol{y}}_{t+h,D} + \tilde{e}_{t+h}$;
(3) **P7:** a constrained regression, $y_{t+h,D} = \alpha + \boldsymbol{w}'_{t+h} \hat{\boldsymbol{y}}_{t+h,D} + \tilde{e}_{t+h}$, s.t $\boldsymbol{w}'_{t+h}\boldsymbol{1} = 1$. We choose to add a constant term, as Granger and Ramanathan (1984) have shown that the fit of the restricted regression without the intercept is not satisfactory. The addition of an intercept helps to control for any forecast biases in the individual forecasts.

*Adaptive elastic net (P8)*

Given that we have many sub-models to consider, it will be interesting to see which forecasting model is most relevant or useful in forecasting the target variable. This can be achieved via regularization or penalized regression. Specifically, instead of using least squares to estimate $\boldsymbol{w}_{t+h}$, we use the adaptive elastic net as described earlier to obtain the weights.

*Random forest (P9)*

We propose using the ensemble-based random forest approach described earlier to non-linearly and nonparametrically aggregate the 16 forecasts. Random forest-based aggregation approaches have been found to outperform both simple and sophisticated combination methods for economic forecasting [46].

*Complete subset regressions (CSR) (P10)*

CSR [47] functions slightly differently from the above weighted aggregations. This approach takes as primitives, not the submodel implied forecasts, but rather the underlying predictors used in the forecasting submodels. In our case, these refer to the exogenous weather and climate variables and their lags, and the cross-disease admission variables (along with their lags). We will focus on estimating Eq. (**2**) but instead of using all the stated predictors, we will extract subsets of the environmental predictors and other categories' ED admissions (and all their lags) and use this subset in the regression (we always include the lags of the target variables). This is done repeatedly with different subsets of predictors. We have 12 environmental variables and 15 other disease causes variables in Eq. (**2**) and we are interested in using subsets with only $p$ variables for each of the two, then there are $(12!/((12-p)!p!)) \times (15!/((15-p)!p!))$ different $p$-variate regression models to estimate. The main goal is to estimate all of the regression models, and compute their implied forecasts. We then combine all the forecasts using a simple average. The challenge here is that depending on either total number of variables or $p$, the number of models to estimate can be large. For example, when $p = 2$, we will have 6930 regressions to consider, and when $p = 3$ we have 100,100. Hence, for computational feasibility, when $p > 1$, we rely on the approach proposed in Elliott et al. (2015) to randomly sample 1000 $p$-variate forecast models without replacement and to average them. We consider $p = 1,2,3$ in our forecasting exercise.

*Random projections (P11)*

Next, we consider a related approach, or an extension called random projections. The framework is similar to (**P10**) in that we are working directly with Eq. (**2**) and taking the predictors here as the primitive. We have two groups of variables here, the environmental variables (including their lags) and the other ED admission variables (including their lags). Let the number of these variables be $k$ and $k'$ respectively. Again, pre-specify a value $p$ (similar to that of P10), and construct the matrices $\boldsymbol{R_k}$ ($k \times p$) and $\boldsymbol{R_{k'}}$ ($k' \times p$) such that each element of both matrices is drawn independently and identically from $N(0,1)$. These matrices will serve as random selection matrices. The forecasting model is then given as

$$y_{t+h,D} = \beta_0 + \sum_{l=0}^{7} \beta_{D,l} y_{t-l,D} + x_t' \boldsymbol{R_k} \beta_{R,k} + y_{t,d \neq D}' \boldsymbol{R_{k'}} \beta_{R,k'} + \tilde{u}_{t+h,D} \tag{18}$$

where $x_t$ contains all the environmental variables and their lags, and $y_{t,d \neq D}$ is a vector with the ED admission data for all the other disease categories and their lags. Here, it is clear that the matrices $\boldsymbol{R_k}$ and $\boldsymbol{R_{k'}}$ will induce a randomly weighted sum of all the predictors in $x_t$ and $y_{t,d \neq D}$. Note that this approach can be seen as a generalization of P10 given that weights other than 0 (not selected) or 1 (selected) can be assigned to the predictors.

Likewise, we vary $p = 1,2,3$, and in each instance, we draw both $\boldsymbol{R_k}$ and $\boldsymbol{R_{k'}}$ randomly 100 times, so we would have 10000 potential regressions. Similar to **P10**, we sample 1000 out of 10000 combinations. Thereafter, we aggregate the forecasts using the simple average of all predictions to generate a final forecast.

For both **P10** and **P11**, we subset-selected or projected environmental covariates and other causes of ED admission separately as associations between environmental variables/other ED case counts with the target disease ED admission category may differ. Doing so would also avoid overfitting and losing any potential spillover effects that different ED categories may have on the target ED admission category.

**Model training and forecast assessment** We subject the constructed data using the sliding window manner for each forecast horizon to an initial 7:3 split, with 70% being training set and 30% being forecast set used for model evaluation. In performing the out-of-sample test, we gradually add the entire forecast set (except the last time point) to the training set using an expanding window approach, we successively update the length of the training set to incorporate as much past information as possible and maximize the model performance. At the first epidemiological week of the forecast period, we generate each respective $h$-week ahead direct forecast with the separate submodels, and train them using the admission-category specific forecasting model described above. For each additional epidemiological week, we include an additional week of observations into the forecasting model fitting. We retain each $h$-week ahead sub-model in order to regenerate the $h$-week ahead conditional forecast.

This expanding window strategy ensures that latest information is incorporated into model estimation and no future data is incorporated in the forecast generated in the contemporaneous time-step for forecasting.

We assess individual submodel and forecast combinations' performance as proxied by their ability to generate accurate 1- to 12-week ahead point forecasts for overall emergency department admissions of each ED category. For point forecasts, actual observations are compared post-hoc against the forecasts of each competing model with forecast performance summarized into 2 error metrics namely: mean absolute percentage error (MAPE), and mean absolute scaled error (MASE).

MAPE summarizes the percentage error the respective forecasting model makes versus the actual observations, and MASE is the mean absolute error of the forecast values, divided by the mean absolute error of the naïve forecast in the initial training set:

$$MAPE = 100 \frac{1}{n} \sum_{t=1}^{n} \left| \frac{y_{t+h,D} - \hat{y}_{t+h,D}}{y_{t+h,D}} \right|^2, \tag{19}$$

$$MASE = \frac{\frac{1}{n}\sum_{t=1}^{n}|y_{t+h,D} - \hat{y}_{t+h,D}|}{\frac{1}{m}\sum_{t=1}^{m}|y_{t+h,D} - y_{t,D}|}, \tag{20}$$

where $n$ and $m$ are the length of the forecast set and initial training set at the $h$-week ahead horizon.

In addition, the one-sided Diebold-Mariano hypothesis test[49] is conducted pairwise between point forecasting models to statistically ascertain the equivalence or non-equivalence at the 5% level.

We further examined whether:

**(1)** the inclusion of a large number of relevant predictors can improve upon forecasts. We predict total ED admissions without and with exogenous covariates included to compare the performance of individual forecasting models in different forecast horizons. Here, we do not include cause-specific ED admission variables as lagged variables, as predictions of total ED admission does not enable us to consider the spillover effect between different causes. As we have changed the number of predictors in this exercise, we only evaluate 10 machine learning models based on regularization, instance, bagging, boosting (except Naïve, HM, AR(A), AR(B), AR(C), and PF).

**(2)** Aggregation of cause-specific ED forecasts to generate forward forecasts of total ED admissions improves upon models which only forecast total ED admissions. To do so, we aggregate cause-specific forecasts at some future horizon to produce forecasts of all-cause ED admissions, and compare this with the forecasts produced by forecasting models trained on the outcome of all-cause total ED admissions.

In both forecasting exercises, the same analytical steps of assessing the forecasts through using train/test splits, expanding window validation, computation of errors and hypotheses tests are done.

## Acknowledgments

This research is funded and supported by the Lee Kong Chian School of Medicine – Ministry of Education Start-Up Grant, by the National Research Foundation Singapore under its Open Fund-Large Collaborative Grant (MOH-001636) and administered by the Singapore Ministry of Health's National Medical Research Council. This research / project is also supported by the Ministry of Education, Singapore, under its Academic Research Fund Tier 1 (RT4/22) and Academic Research Fund Tier 1 Seed Funding Grant (RS04/22).

## Author contributions

Conceptualization: J.T.L., K.M., P.G. Methodology: J.T.L., K.M., P.G., W.Y.L. Software: P.G., W.Y.L. Validation: P.G., W.Y.L., J.T.L., K.M. Formal Analysis: P.G., W.Y.L., J.T.L., K.M. Investigation: P.G., W.Y.L., J.T.L., K.M. Resources: J.T.L., B.L.D. Data Curation: J.T.L., B.L.D., K.B.T., J.A., P.M. Writing—original draft: P.G., W.Y.L., J.T.L., K.M. Writing—review & editing: P.G., W.Y.L., J.T.L., K.M., E.L.W.C, B.L.D., K.B.T., J.A., P.M. Visualization: P.G., W.Y.L. Funding acquisition: J.T.L. Supervision: J.T.L., K.M. Project Administration: J.T.L.

## Conflict of Interests

The authors declare no conflict of interests.

## Data availability

The data that support the findings of this study are available from Ministry of Health, Singapore, but restrictions apply to the availability of these data, which were used under licence for the current study, and so are not publicly available. Data are however available from the authors upon reasonable request and with permission of Ministry of Health, Singapore.

## Code availability

The code implementation of this study is made publicly available at https://github.com/gpeihong/lkc-point-forecast

## Supplementary material

S1 Text. Appendix containing additional details on data, methods and results.

**S2 File.zip** Plots of case counts versus forecasts.

# High-dimensional point forecast combinations for emergency department demand

Supplementary Material (S1 Text)

**Table 1 in S1 Text: Descriptive statistics and data sources for predictive variables**

| Predictors | Mean | Standard deviation | Max | Min | Source |
|---|---|---|---|---|---|
| Cardiovascular disease | 842.129 | 75.705 | 1075 | 601 | Ministry of Health, Singapore |
| Chronic respiratory disease | 586.518 | 71.922 | 908 | 412 | Ministry of Health, Singapore |
| Diabetes mellitus | 70.931 | 14.04 | 114 | 26 | Ministry of Health, Singapore |
| Digestive disease | 1297.075 | 258.093 | 1852 | 648 | Ministry of Health, Singapore |
| Endocrine disorders | 279.042 | 65.639 | 466 | 160 | Ministry of Health, Singapore |
| Factors influencing health status and contact with health services | 98.294 | 21.877 | 176 | 52 | Ministry of Health, Singapore |
| Genitourinary disorders | 638.219 | 81.138 | 835 | 415 | Ministry of Health, Singapore |
| Ill-defined diseases | 3424.891 | 449.697 | 4494 | 2338 | Ministry of Health, Singapore |
| Infectious and Parasitic Diseases | 1301.681 | 217.162 | 2029 | 844 | Ministry of Health, Singapore |
| Malignant neoplasms | 77.752 | 11.341 | 115 | 41 | Ministry of Health, Singapore |
| Musculoskeletal disease | 1041.985 | 219.707 | 1496 | 582 | Ministry of Health, Singapore |
| Neurological and sense disorders | 798.942 | 79.232 | 1015 | 543 | Ministry of Health, Singapore |
| Oral diseases | 69.075 | 12.37 | 128 | 39 | Ministry of Health, Singapore |
| Other neoplasms | 16.685 | 4.646 | 32 | 5 | Ministry of Health, Singapore |
| Respiratory infection | 2511.987 | 608.874 | 6649 | 1266 | Ministry of Health, Singapore |
| Skin diseases | 713.522 | 75.591 | 899 | 490 | Ministry of Health, Singapore |
| Maximum temperature (K) | 304.865 | 0.979 | 307.458 | 300.979 | ERA5-Land |
| Mean temperature (K) | 300.982 | 0.859 | 303.21 | 298.095 | ERA5-Land |
| Minimum temperature (K) | 298.116 | 0.801 | 300.773 | 296.107 | ERA5-Land |
| Relative humidity (%) | 79.464 | 3.898 | 91.475 | 63.546 | ERA5-Land |
| Absolute humidity (g/m$^3$) | 21.33 | 0.878 | 23.463 | 17.315 | ERA5-Land |
| Total precipitation (m) | 0.005 | 0.006 | 0.035 | 0 | ERA5-Land |
| PM2.5 (μg/m$^3$) | 18.367 | 11.063 | 167.028 | 9.164 | National Environment Agency, Singapore |
| PM10 (μg/m$^3$) | 29.965 | 13.491 | 213.948 | 16.664 | National Environment Agency, Singapore |
| O3 (μg/m$^3$) | 24.299 | 7.969 | 59.566 | 10.036 | National Environment Agency, Singapore |
| NO2 (μg/m$^3$) | 23.863 | 5.703 | 44.317 | 10.342 | National Environment Agency, Singapore |
| SO2 (μg/m$^3$) | 10.78 | 5.025 | 30.238 | 2.16 | National Environment Agency, Singapore |
| CO (mg/m$^3$) | 0.537 | 0.132 | 2.152 | 0.296 | National Environment Agency, Singapore |

**Table 2 in S1 Text: Submodels and forecast combinations used for emergency department disease forecasting**

| Model | Model Type |
|---|---|
| Naïve | Baseline |
| Historical mean | Simple |
| Autoregression without exogenous variables | Simple |
| Autoregression with environmental variables | Simple |
| Autoregression with all exogenous variables | Simple |
| Ridge Regression | Simple |
| LASSO | Simple |
| Adaptive LASSO | Simple |
| Sparse Group LASSO | Simple |
| Elastic Net | Simple |
| Adaptive Elastic Net | Simple |
| Pure Factor Model | Simple |
| Random Forest | Simple |
| K-Nearest Neighbours | Simple |
| XGBoost | Simple |
| LightGBM | Simple |
| Mean | Simple Combinations |
| Median | Simple Combinations |
| Bates and Granger | Simple Combinations |
| Bates and Granger (Weighted-Mean with Expanding Window) | Simple Combinations |
| 2nd-Order Linear Regression | Supervised Learning Combinations |
| 2nd-Order Linear Regression Omitting Intercept | Supervised Learning Combinations |
| 2nd-Order Constrained Linear Regression | Supervised Learning Combinations |
| 2nd-Order Adaptive Elastic Net | Supervised Learning Combinations |
| 2nd-Order Random Forest | Supervised Learning Combinations |
| Complete Subset Regressions | Supervised Learning Combinations |
| Random Projections | Supervised Learning Combinations |

**Figure 1 in S1 Text: Forecast assessment statistics for the all-cause and cause-specific forecasting models in the full forecast period.**

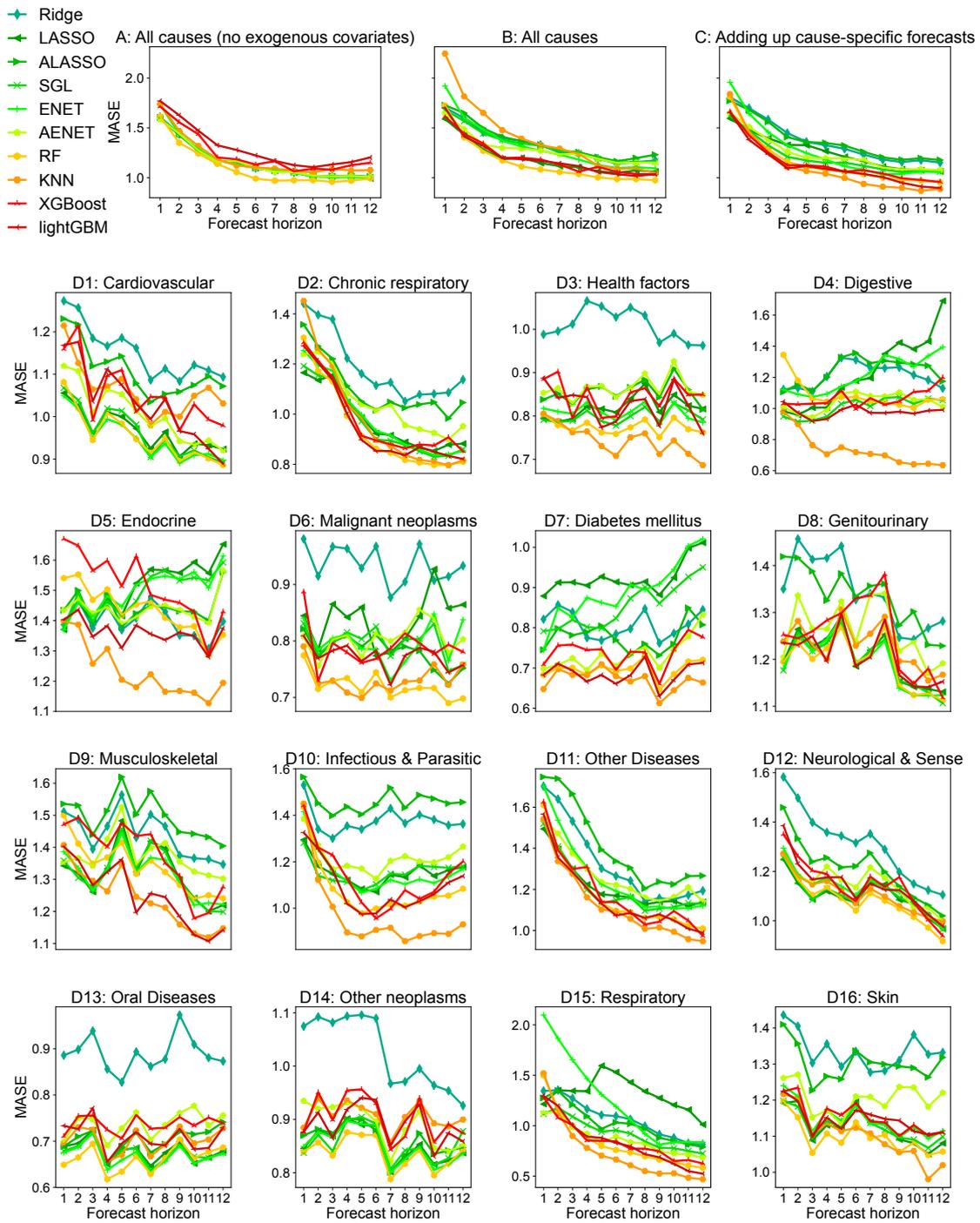

Mean absolute scaled error (MASE) of the 10 individual simple forecasting models which include: ridge regression (Ridge), Least Absolute Shrinkage and Selection Operator (LASSO), adaptive LASSO (ALASSO), sparse group LASSO (SGL), elastic net (ENET), adaptive elastic net (AENET), random forest (RF), k-nearest neighbors (KNN), XGBoost and lightGBMs, across the forecast horizon of 1-12 weeks ahead. (A) value assessed: forecasts of the total number of ED admissions for all causes; model inputs: autoregressive term (B) value assessed: forecasts of the total number of ED admissions for all causes; model inputs: autoregressive term and environmental covariates. (C) value assessed: sum of forecasts of the casue-specific ED admissions; model inputs: autoregressive term and environmental covariates. (D1–D16) value assessed: forecasts of the casue-specific ED admissions; model inputs: autoregressive term and environmental covariates.

**Figure 2 in S1 Text: Contributions of individual submodels in the weights-based forecast combinations for 4 disease categories.**

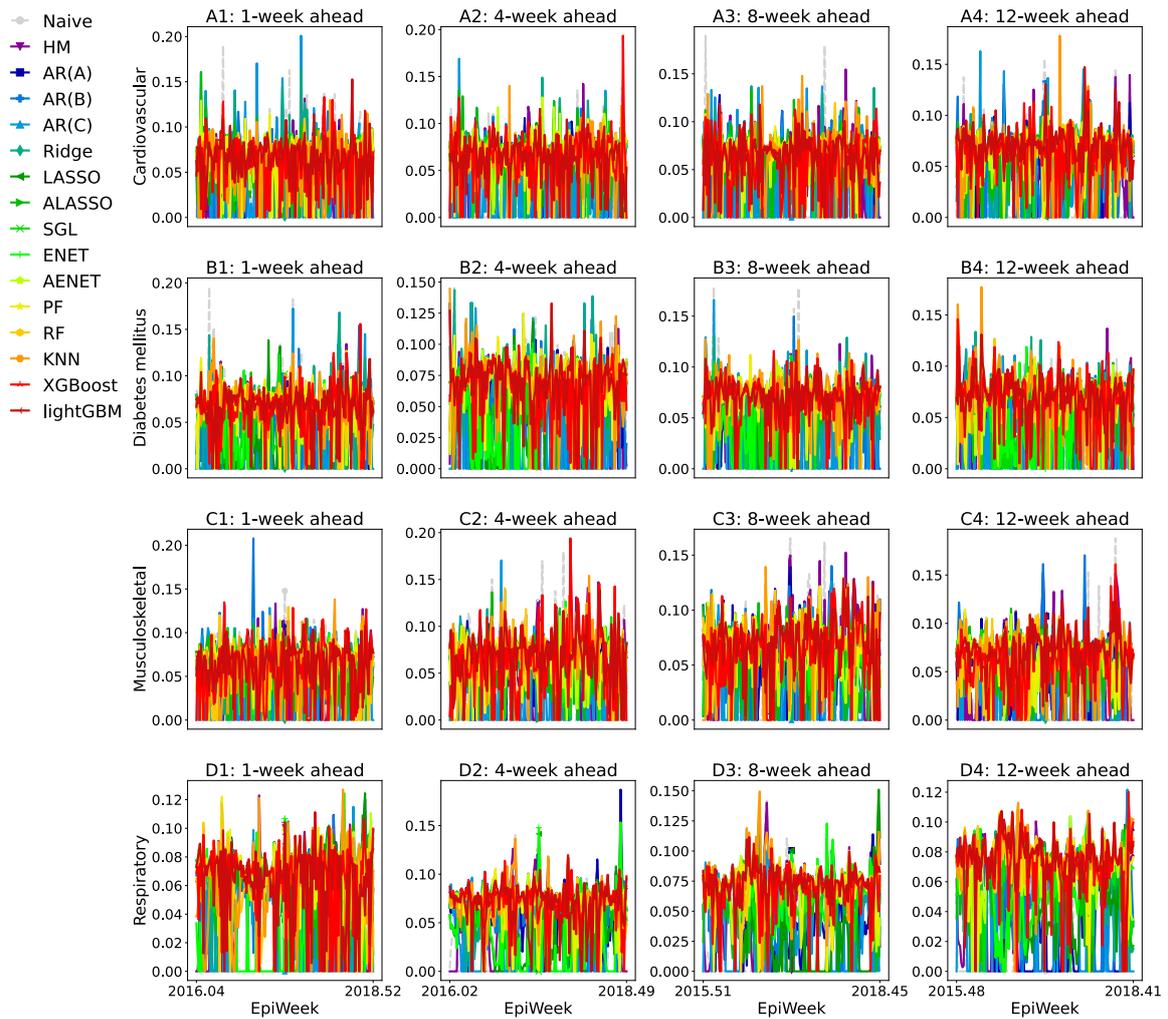

Visualization of how the weights of individual submodels in the contemporaneous time-based Bates and Granger (P3) change over epidemiological weeks for 4 disease categories: Cardiovascular disease, Diabetes mellitus, Musculoskeletal disease, Respiratory Infection and 4 forecasting windows (1-week, 4-week, 8-week, 12-week).

**Figure 3 in S1 Text: Contributions of individual submodels in the weights-based forecast combinations for 4 disease categories.**

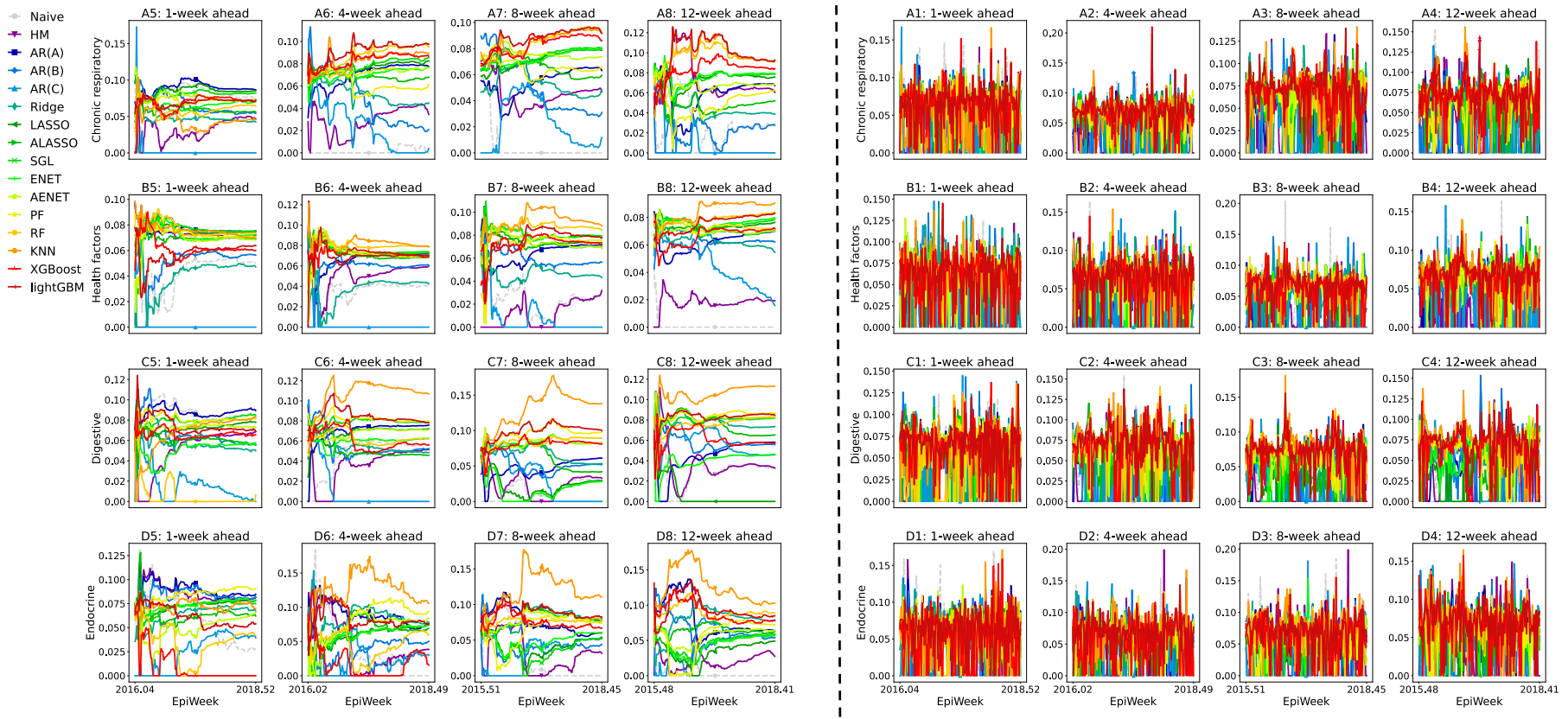

Visualization of how the weights of individual submodels in the expanding window-based Bates and Granger (P4) (left of the dotted line) and the contemporaneous time-based Bates and Granger (P3) (right of the dotted line) change over epidemiological weeks for 4 disease categories: Chronic respiratory disease, Factors influencing health status and contact with health services, Digestive disease, Endocrine disorders and 4 forecasting windows (1-week, 4-week, 8-week, 12-week).

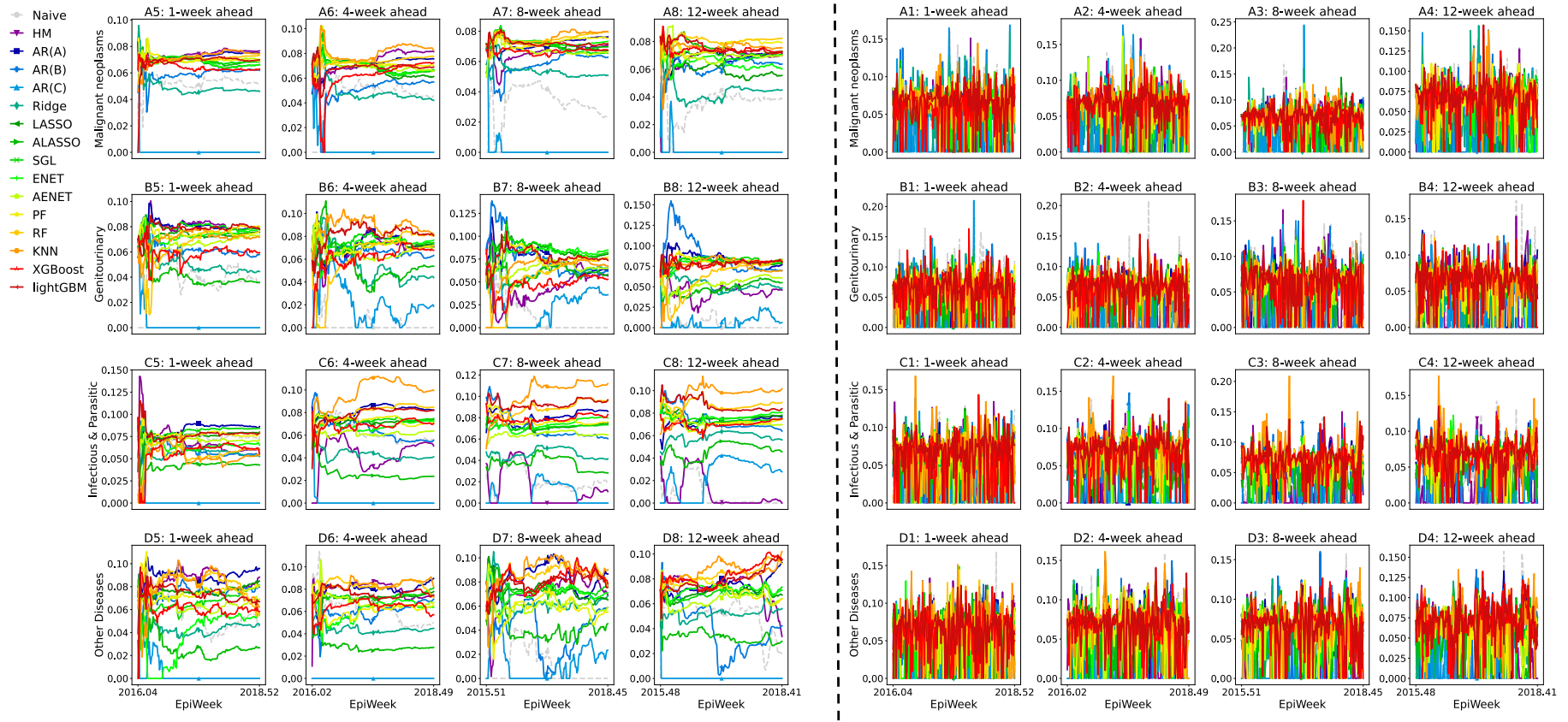

**Figure 4 in S1 Text: Contributions of individual submodels in the weights-based forecast combinations for 4 disease categories.**

Visualization of how the weights of individual submodels in the expanding window-based Bates and Granger (P4) (left of the dotted line) and the contemporaneous time-based Bates and Granger (P3) (right of the dotted line) change over epidemiological weeks for 4 disease categories: Malignant neoplasms, Genitourinary disorders, Infectious and Parasitic Diseases, Ill-defined diseases and 4 forecasting windows (1-week, 4-week, 8-week, 12-week).

**Figure 5 in S1 Text: Contributions of individual submodels in the weights-based forecast combinations for 4 disease categories.**

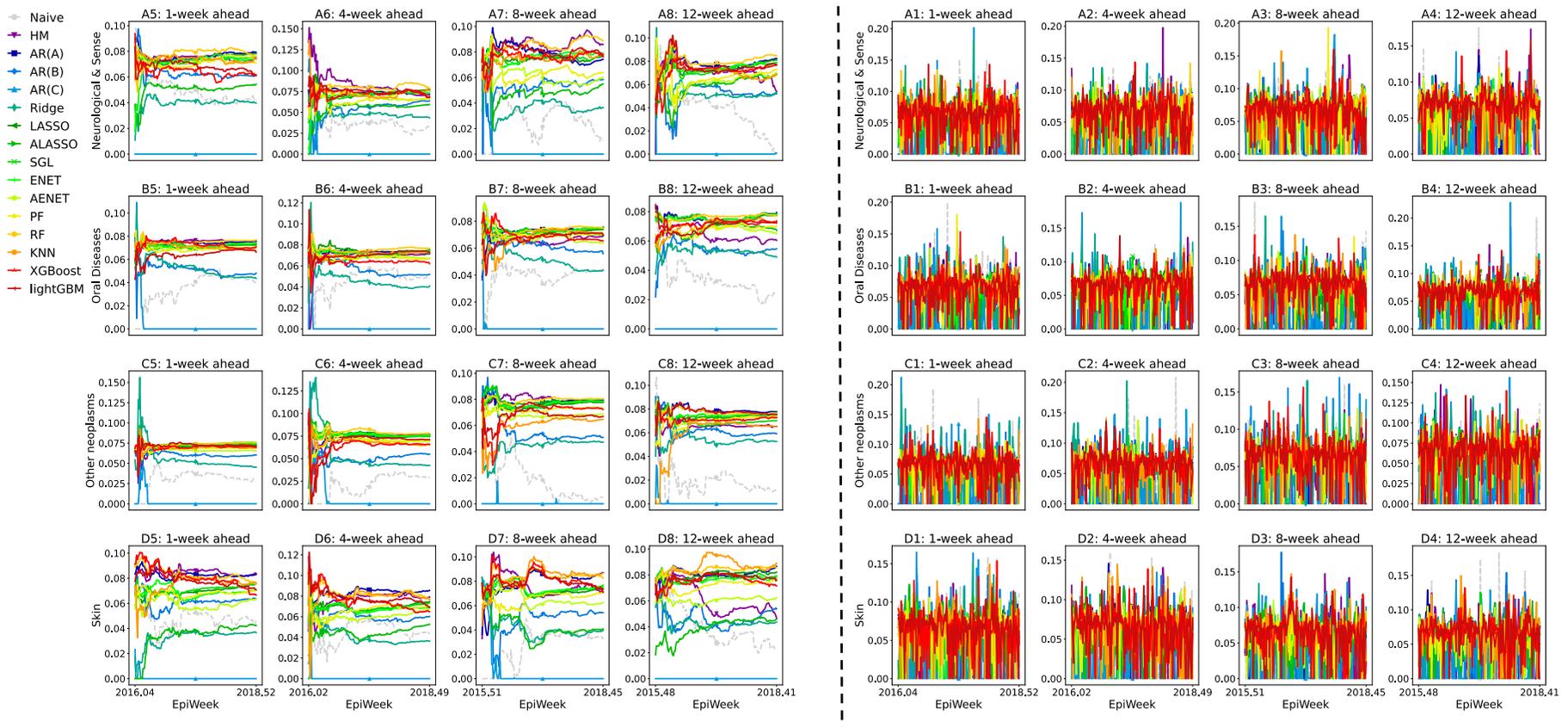

Visualization of how the weights of individual submodels in the expanding window-based Bates and Granger (P4) (left of the dotted line) and the contemporaneous time-based Bates and Granger (P3) (right of the dotted line) change over epidemiological weeks for 4 disease categories: Neurological and sense disorders, Oral Diseases, Other neoplasms, Skin diseases and 4 forecasting windows (1-week, 4-week, 8-week, 12-week).